\documentclass[twocolumn, superscriptaddress]{revtex4-2} 
\usepackage[utf8]{inputenc}
\usepackage[a4paper, left=15mm, right=15mm, top=25mm, bottom=25mm]{geometry}
\usepackage{graphicx}
\usepackage{siunitx}
\usepackage{amssymb}
\usepackage{amsmath}
\usepackage{appendix}
\usepackage[dvipsnames]{xcolor}
\usepackage[normalem]{ulem}
\usepackage{braket}
\DeclareSIUnit\gauss{G}

\begin{document}

%\title{Collision energy dependence of atom-ion Feshbach resonances below the $s$-wave limit}
\title{Exploring atom-ion Feshbach resonances below the $s$-wave limit}

\author{Fabian Thielemann}
    \affiliation{Physikalisches Institut, Albert-Ludwigs-Universität Freiburg, Hermann-Herder Str. 3, 79104
    Freiburg, Germany}
\author{Joachim Siemund}
    \affiliation{Physikalisches Institut, Albert-Ludwigs-Universität Freiburg, Hermann-Herder Str. 3, 79104
    Freiburg, Germany}
\author{Daniel von Schoenfeld}
    \affiliation{Physikalisches Institut, Albert-Ludwigs-Universität Freiburg, Hermann-Herder Str. 3, 79104
    Freiburg, Germany}
\author{Wei Wu}
    \affiliation{Physikalisches Institut, Albert-Ludwigs-Universität Freiburg, Hermann-Herder Str. 3, 79104
    Freiburg, Germany}
    \affiliation{EUCOR Centre for Quantum Science and Quantum Computing, Albert-Ludwigs-Universität Freiburg,
    79104 Freiburg, Germany}
\author{Pascal Weckesser}
    \affiliation{Physikalisches Institut, Albert-Ludwigs-Universität Freiburg, Hermann-Herder Str. 3, 79104
    Freiburg, Germany}
    \affiliation{Max-Planck-Institut f\"{u}r Quantenoptik, 85748 Garching, Germany}
    \affiliation{Munich Center for Quantum Science and Technology (MCQST), 80799 Munich, Germany}
\author{Krzysztof Jachymski}
    \affiliation{Faculty of Physics, University of Warsaw, Pasteura 5, 02-093 Warsaw, Poland}
\author{Thomas Walker}
    \affiliation{Physikalisches Institut, Albert-Ludwigs-Universität Freiburg, Hermann-Herder Str. 3, 79104
    Freiburg, Germany}
    \affiliation{EUCOR Centre for Quantum Science and Quantum Computing, Albert-Ludwigs-Universität Freiburg,
    79104 Freiburg, Germany}
    \affiliation{Blackett Larboratory, Imperial College London, Prince Consort Road, London SW7 2AZ, United
    Kingdom}
\author{Tobias Schaetz}
    \affiliation{Physikalisches Institut, Albert-Ludwigs-Universität Freiburg, Hermann-Herder Str. 3, 79104
    Freiburg, Germany}
    \affiliation{EUCOR Centre for Quantum Science and Quantum Computing, Albert-Ludwigs-Universität Freiburg,
    79104 Freiburg, Germany}

\date{\today}

\begin{abstract}
    Hybrid systems of single, trapped ions, embedded in quantum gases are a promising platform for quantum simulations and the study of long-range interactions in the ultracold regime.
    Feshbach resonances allow for experimental control over the character and strength of the atom-ion interaction.
    However, the complexity of atom-ion Feshbach spectra, e.g. due to second-order spin-orbit coupling, requires a detailed experimental understanding of the resonance properties – such as the contributing open-channel partial waves.
    In this work, we immerse a single barium (Ba$^+$) ion in a bath of lithium (Li) atoms spin-polarized in their hyperfine ground state to investigate the collision energy dependence of magnetically tunable atom-ion Feshbach resonances.
    We demonstrate fine control over the kinetic energy of the Ba$^+$ ion and employ it to explore three-body recombination in the transition from the many- to the few-partial wave regime, marked by a sudden increase of resonant loss.
    In a dense spectrum – with on average 0.58(1) resonances per Gauss – we select a narrow, isolated feature
    and characterize it as an $s$-wave resonance. We introduce a quantum recombination model that allows us to distinguish it from higher-partial-wave resonances.
    Further, in a magnetic field range with no significant loss at the lowest collision energies, we identify
    an higher-partial-wave resonance that only appears and peaks when we increase the energy to around the $s$-wave limit.
    Our results demonstrate that hybrid atom-ion traps can reach collision energies well in the ultracold regime, and that the 
    ion's kinetic energy can be employed to tune the collisional complex to resonance, paving the way for fast control over
    the interaction in settings where magnetic field variations are detrimental to coherence.
\end{abstract}
\maketitle

\newpage

\section{Introduction}
%\begin{itemize}
    %\item 
Resonant scattering is of major importance in a variety of physical processes, ranging from particle creation to photons interacting with optical cavities.
    %\item 
At the lowest temperatures, where particles exhibit wave-like behavior, interference effects can drastically
        alter the outcome of a collision \cite{dalibardCollisionalDynamicsUltracold1999}.
    %\item 
    The transition into this regime is typically marked by reaching collision energies below the $s$-wave limit; that
    is the
    energetic height of the lowest collisional angular momentum barrier ($\ell=1$).
    For most neutral gases the $s$-wave limit is well above the Doppler temperature, facilitating the access
    to Feshbach resonances that fuel the ongoing investigation of various many-particle
    Hamiltonians or the interaction between atoms at close
    range \cite{chinFeshbachResonancesUltracold2010,blochManybodyPhysicsUltracold2008b}.
    %\item 
    Near a Feshbach resonance, external magnetic fields can be used to vary the
atomic interaction from attractive to repulsive or even tune it out entirely -- experimentally evidenced by
monitoring the loss of colliding particles from the trap.
    %\item 
%
Studying Feshbach resonances in various energy regimes can reveal the physical laws behind the involved loss processes. 
This has been
used to demonstrate how Pauli's exclusion principle suppresses the $s$-wave scattering of ultracold
spin-polarized fermions \cite{demarcoMeasurementWaveThreshold1999}, to understand the chaotic Feshbach resonance spectra
of Lanthanides \cite{maierEmergenceChaoticScattering2015}, or to study a novel resonant loss process in
ultracold molecular collisions \cite{parkFeshbachResonanceCollisions2023}. These studies of neutral
gases are typically constrained to energies below the $p$-wave barrier by the finite depth of optical traps. On the other hand, merged beam experiments
with precise control over collision energies of a few millikelvin and above have unveiled intricate
details of quantum resonant loss processes like the role of Feshbach resonance pathways on the final state
distribution \cite{margulisTomographyFeshbachResonance2023, hensonObservationResonancesPenning2012}.
Applying similar methods to novel platforms, such as atom-ion systems in the ultracold regime, is particularly interesting.

\begin{figure*}[t]
    \centering
\includegraphics{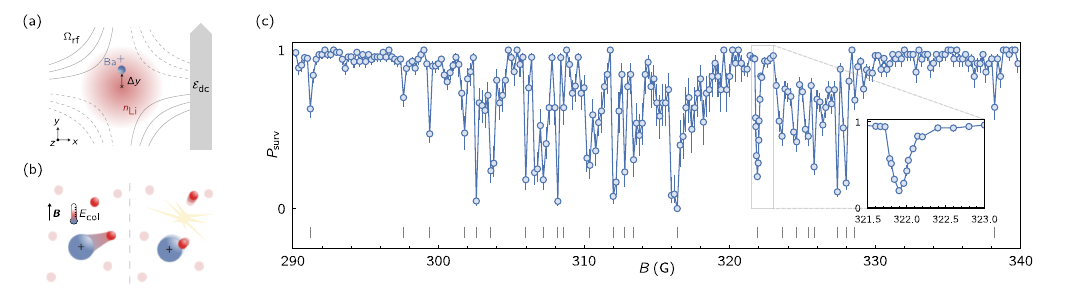}
\caption{\textbf{Experimental control over atom-ion scattering via Feshbach resonances}\\
    (a)
    %\begin{itemize}
        %\item 
Transverse cut through our hybrid trap setup.
        %\item 
A single Ba$^+$ ion is confined in a linear radio frequency quadrupole trap (dashed and solid lines) and embedded in a cloud of ultracold, spin-polarized fermionic $^6$Li (red) near degeneracy.
        %\item 
An electric field $\mathcal{E}_\text{dc}$ (gray arrow) is applied to displace the ion from the center of the rf trap and
control the excess kinetic energy of the ion.
    %\end{itemize}
    (b)
    %\begin{itemize}
        %\item 
Illustration of a two-step three-body recombination process:
        %\item 
The Ba$^+$ ion is interacting with two $^6\text{Li}$ atoms. 
        %\item 
Depending on the magnetic field $B$ and the collision energy, a resonant LiBa$^+$ dimer state may be formed. 
        %\item 
Next, the dimer deexcites to a deeply bound molecular state, releasing its binding energy in the form
            of kinetic energy of the collision complex.
    %\end{itemize}
        (c) 
    %\begin{itemize}
        %\item 
        The ion-loss spectrum between 290 and \SI{340}{\gauss} (blue circles), recorded at lowest collision energy, 
            reveals 24 Feshbach resonances (vertical dashes).
        %\item 
        Around \SI{322}{\gauss} (inset) we find a narrow, but pronounced, resonance. The connecting lines
        serve as a guide to the eye. Errorbars indicate $1\sigma$ confidence intervals.
    %\end{itemize}
    } 
    \label{fig:fig1}

\end{figure*}
An ion, embedded in an ultracold atomic gas, is an intriguing system to study collision energy effects at
ultracold temperatures: the ion serves as a single,
highly controllable probe and interacts with the atoms via the long-range isotropic charge-induced-dipole
interaction, allowing novel applications ranging from quantum simulations to cold chemistry \cite{harterColdAtomIon2014,
tomzaColdHybridIonatom2019, deissColdTrappedMolecular2024, karmanUltracoldChemistryTestbed2024}. However, the
$1/R^4$ scaling (where $R$ is the distance between ion and atom) at long
        range implies that much lower collision energies are required to enter the $s$-wave regime. As an
        example, the $s$-wave limit for Li-Ba$^+$ is
        $E_s = \SI{8.8}{\micro\kelvin}$, in contrast to the orders of magnitude larger $E_{s,\text{Li-Li}}
        \approx \SI{8}{\milli\kelvin}$ in the case of Li-Li collisions, a workhorse in the field of ultracold
        atom experiments.
    %\item 
In fact, the few-partial wave regime has only recently been reached, as experimentally
witnessed by a variation of the spin-exchange rate in $^6$Li-$^{171}$Yb$^+$ and the direct observation of Feshbach
        resonances in $^6$Li-$^{138}$Ba$^+$
        \cite{feldkerBufferGasCooling2020b, weckesserObservationFeshbachResonances2021}.
    %\item 
%    
In the latter case with Li polarized in the second lowest hyperfine state, 13 resonances were
discovered within a range of approximately \SI{100}{\gauss},
        substantially more than in most alkali atom-atom combinations. 
        The unusually high density of Feshbach resonances is in part due to overlap between the
        singlet and triplet ground states of the LiBa$^+$ molecule with its electronically
        excited $b^3\Pi$ state that lead to significant second-order spin-orbit interaction and
        allow coupling between different total spin channels. This additional coupling turned out to add
        considerable complexity to numerical calculations 
        \cite{tscherbulSpinOrbitInteractionsQuantum2016, weckesserObservationFeshbachResonances2021}.
    %\item 
    An experimental investigation of the energy dependence of individual Feshbach resonances could provide the missing
    ingredient for numerical calculations.
    So far, studying the energy dependence of inelastic atom-ion collisions in the many-partial-wave regime shed light on three-body recombination 
    processes \cite{krukowEnergyScalingCold2016, perez-riosCommunicationClassicalThreshold2015}.
    Extending this control to energies below $E_s$ could reveal the microscopic properties of atom-ion Feshbach 
    resonances, such as their partial-wave assignment, and allow for a new level of control over complex many-body dynamics at ultracold temperatures.

In this Article, we demonstrate collision energy tuning below the atom-ion $s$-wave limit $E_s$ and apply our
method to assign the partial-wave order of selected Feshbach resonances.
Preparing the atoms in their hyperfine ground state and the ion at lowest kinetic energy, we find a dense 
Feshbach spectrum with more than 40 resonances in a range of \SI{100}{\gauss}. 
We use the collision energy tuneability to vary the atom-ion collision energy over several orders of magnitude near one of the resonances and
observe the transition from the many-partial-wave to the $s$-wave regime, witnessed by a sharp modulation of
ion loss.
Examining the resonance more closely at energies below $E_s$, we find a strong dependence of its amplitude
on the collision energy. The behavior can be described by modeling an $s$-wave resonance with a beyond threshold quantum
recombination model.
Additionally, in a nearby magnetic field region, where we detect no resonance at the lowest collision
energies, we observe a resonance that appears and peaks at energies around and above $E_s$.
We find qualitative agreement between this observation and a theoretically modeled $f$-wave resonance,
underlining the importance of taking higher-partial-wave contributions into account, even at collision
energies below $E_s$.

\section{Experimental setup and ion loss spectroscopy}
%\begin{itemize}
        We use a hybrid trap setup consisting of a radio-frequency (rf) trap for a single $^{138}$Ba$^+$ ion and a crossed optical dipole
        trap (xODT) for fermionic $^6$Li atoms \cite{weckesserTrappingShapingIsolating2021,
        weckesserObservationFeshbachResonances2021} (see Fig.~\ref{fig:fig1}a).
        The atoms are spin-polarized in their lowest lying hyperfine state $\ket{1}_\text{Li}$ and cooled to a temperature of
        $T_\text{Li}\approx \SI{700(50)}{\nano\kelvin}$.
        We prepare the ion either in its electronic S$_{1/2}$ ground or D$_{3/2}$ metastable excited state.
        We allow it to interact with the atoms for $t_\text{int}$ and monitor its survival probability
        $P_\text{surv}$ (see Appendix~\ref{apdx:ion_state}).
        During the interaction, we apply an external magnetic field $B$ and, optionally, an electric displacement
        field $\mathcal{E}_\text{dc}$.
    %\item

\begin{figure}[t]
    \centering
    \includegraphics[]{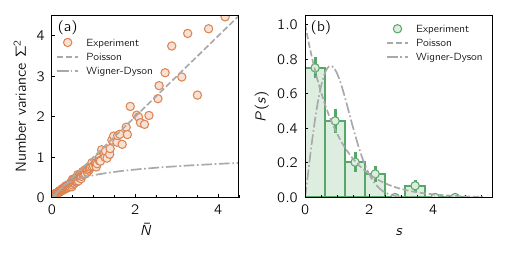}
    \caption{\textbf{Statistical analysis of the resonance spectrum:} (a) The variance
    $\Sigma^2$ of the number of resonances in an interval is plotted as a function of
    the average number of resonances in that interval $\bar{N}$. The experimental data (orange markers) are
    consistent with the scaling of Poisson-distributed resonances (dashed line) and deviate significantly from
    a Wigner-Dyson distribution. (b) The
    probability distribution of the normalized nearest-neighbor spacing $s=\overline{\rho}\,\delta B$ shows consistency 
    between the resonance positions observed experimentally (green markers) and a Poisson-distributed
    sample (dashed line). For comparison, a Wigner-Dyson distribution, featuring the characteristic
    anti-bunching, is also plotted (dash-dotted line). The errorbars indicate 1$\sigma$ confidence intervals.}
    \label{fig:resonance_stats}
\end{figure}

        The combination of ion and atoms in their respective electronic ground state is chemically stable
        \cite{tomzaColdHybridIonatom2019}.
        Two-body charge exchange collisions, which can occur in other species combinations, are
        endothermic. This means
        that three-body recombination (TBR) is the dominant process for ion loss
        \cite{weckesserObservationFeshbachResonances2021} (Fig.~\ref{fig:fig1}b).
    %\item 
Measuring the dependence of TBR on $B$ reveals the atom-ion Feshbach spectrum, a part of which is
shown in Fig.~\ref{fig:fig1}c.
    %\item 
In a dense spectrum, we observe a narrow atom-ion Feshbach resonance at \SI{321.90(3)}{\gauss} with a full width at half
        maximum (FWHM) of \SI{250(20)}{\milli\gauss}, clearly distinguishable from the surrounding background.
        This makes this resonance an ideal candidate to study collision energy effects and a good
        reference for the stability of $B$.
          In total, we sample a range of \SI{100}{\gauss} with individual scans of
        \SI{10}{\gauss} with step size of \SI{200}{\milli\gauss}. Interleaved with this we sample the
        \SI{321.90(3)}{\gauss} resonance with a finer step size to rule out magnetic field drifts or loss of
        contrast due to a decrease in overlap between the ion and the atoms.
    %\item 
We classify individual resonances as local minima that are separated from the next local minimum by a barrier
        of at least $3\sigma$ height. 
        In this way, we identify 49 resonances of different widths and
        amplitudes in the range of \SIrange{240}{340}{\gauss}, on average
        $\overline{\rho}=\SI{0.58(1)}{\per\gauss}$.
    %\item 
In the spectra of lanthanides that exhibit a high resonance density (e.g. \SI{3.4}{\per\gauss} for $^{168}$Er 
at comparable collision energy~\cite{maierEmergenceChaoticScattering2015}), a statistical analysis revealed properties of chaotic
        scattering.
        While strongly correlated energy levels would result in chaotic scattering and a spectrum that follows Wigner-Dyson
        statistics, non-interacting levels give a Poissonian distribution.
    %\item 
We follow the analysis of the number variance $\Sigma^2$ and normalized nearest neighbor spacing
$s=\overline{\rho}\, \delta B$ as
    presented in \cite{maierEmergenceChaoticScattering2015, frischQuantumChaosUltracold2014a,
        khlebnikovRandomChaoticStatistic2019} and show that both are in good agreement with a Poissonian (i.e.
        non-interacting) distribution of resonances (see Fig.~\ref{fig:resonance_stats}). Both $\Sigma^2$ and $P(s)$ deviate significantly from a
        Wigner-Dyson distribution that would be expected for chaotic scattering. 
        The authors of \cite{naubereitIntrinsicQuantumChaos2018} report that other factors, besides the nature of
        the underlying scattering dynamics, can influence the distribution of the measured resonances. Both, the
        superposition of spectra (such as those belonging to the two Zeeman states of Ba$^+$), and
        undetected resonances due to missing experimental resolution can lead to Poissonian statistics.
        However, we estimate that about \SI{90}{\percent} of resonances would have to be undetected to obtain 
        a distribution as close to Poissonian as what we observe. To within our experimental resolution, we
        observe no signature of chaotic scattering. This would be in agreement with the recent
        findings of \cite{pinkasChaoticScatteringUltracold2024} who give an expression for the critical
        kinetic energy of the atoms,
        below which chaotic scattering dynamics could be expected. It evaluates to
        $E_a^c\approx\SI{35}{\nano\kelvin}\times k_\text{B}$ for our parameters, significantly lower than
        $T_\text{Li}$.
    %\item 
%

        Conversely, preparation of the ion in the metastable D$_{3/2}$ state allows for additional inelastic two-body loss
        processes. At our typical densities, the corresponding ion-loss rate $\gamma_\text{l}$ is at least an order of magnitude
        higher than that of TBR \cite{xingCompetingExcitationQuenching2024a}. As a two-body process,
          $\gamma_\text{l}$ depends linearly on the atomic density 
          $ \gamma_\text{l} \propto n$, allowing us to probe the profile of the atomic cloud \cite{jogerObservationCollisionsCold2017a}.
          As it is based on Langevin collisions, $\gamma_l$ is not expected
          to show any dependence on the collision energy \cite{tomzaColdHybridIonatom2019}.
%\end{itemize}

\section{Controlling the collision energy}

\begin{figure}[t]
    \centering
    \includegraphics[]{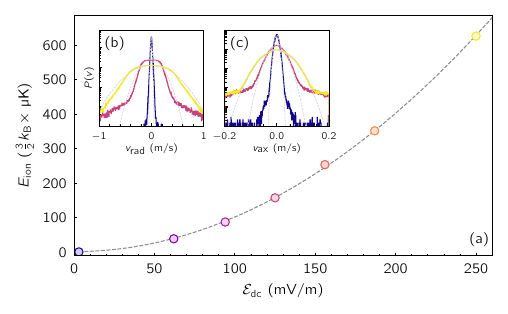}
    \caption{\textbf{Ion kinetic energy scaling:} (a) The lab frame median ion kinetic energy $E_\text{ion}$, based on our
        simulations, is shown in dependence on the applied displacement field $\mathcal{E}_\text{dc}$ (circles). The
    data are fitted with a quadratic function to obtain $\alpha_\text{sim}$ (gray dashed line). (b) and (c) The 
    radial and axial velocity distribution for $3(125/250)\si{\milli\volt\per\meter}$ (blue/purple/yellow
    line) are shown in a semi-logarithmic scale. The grey dashed lines are fits with a Gaussian (thermal)
    velocity distribution. At low $\mathcal{E}_\text{dc}$ it is a good description in both the radial and the axial
    degree of freedom. At increased $\mathcal{E}_\text{dc}$ the radial velocity distribution shows stronger
    deviations due to the direct effect of rf driven micromotion. The axial degree of freedom is only heated
    by collisional redistribution of the radial kinetic energy and shows a more thermal behavior.}
    \label{fig:energy_scaling}
\end{figure}
%\begin{itemize} 
    %\item 
The application of a radial displacement field $\mathcal{E}_\text{dc}$ during $t_\text{int}$ increases the
kinetic energy of the ion $E_\text{ion}$. This is due
        to the increase of excess micromotion when the ion is displaced from the rf null \cite{berkelandMinimizationIonMicromotion1998,
        cetinaMicromotionInducedLimitAtomIon2012, krukowEnergyScalingCold2016}. The motion is driven at
            the rf drive frequency $\Omega_\text{rf}$ and, combined
        with elastic collisions with the atoms, results in a steady-state ion kinetic energy
        distribution that is non-thermal \cite{pinkasEffectIontrapParameters2020a,
    holtkemeierBufferGasCoolingSingle2016}. In the following, we denote the resulting excess median kinetic
    energy of the ion as $\Delta E_\text{ion}$. Over the range of displacement fields $\mathcal{E}_\text{dc}$ explored here, $\Delta
    E_\text{ion}$ scales quadratically 
    \begin{align}
        E_\text{ion} &= E_\text{min} + \Delta E_\text{ion} \notag\\
                     &= E_\text{min} + \alpha \mathcal{E}_\text{dc}^2.
    \end{align}
        Using molecular dynamics (MD) simulations, we determine the proportionality of
        $\alpha_\text{sim} =
        \SI{10(1)}{\milli\kelvin/(\volt\per\meter)^{2}}$ in our setup (Fig.~\ref{fig:energy_scaling}).
        In this way, we can experimentally fine-tune the collision energy
        by applying $\mathcal{E}_\text{dc}$.
    %\item 
Note that the energy scale relevant for collisions is defined in the center of mass (COM) and
        can be more than twenty
        times lower than $\Delta E_\text{ion}$ due to
        the mass imbalance of $m_\text{Li}/m_\text{Ba} = 6/138$ and the ultracold temperature of the atomic bath.

    %\item 
The transition to the $s$-wave regime, determined by a COM collision energy below $E_s$, is reached
        for $\Delta E_{s,\text{ion}}\approx
        \SI{195}{\micro\kelvin}\times \frac{3}{2} k_\text{B}$, for an atomic bath temperature of \SI{700(50)}{\nano\kelvin}. 
    %\item 
Our simulations suggest that the ion, sympathetically cooled by the atomic bath, reaches an equilibrium median
kinetic energy of $E_\text{min} = \SI{2.2(2)}{\micro\kelvin}\times
        \frac{3}{2}k_\text{B}$ at stray electric fields compensated to within $\mathcal{E}_\text{dc}\approx
        \SI{3}{\milli\volt\per\meter}$, our current experimental accuracy.

\section{From the many-partial-wave to the $s$-wave regime}

\begin{figure}[t]
    \centering
    \includegraphics{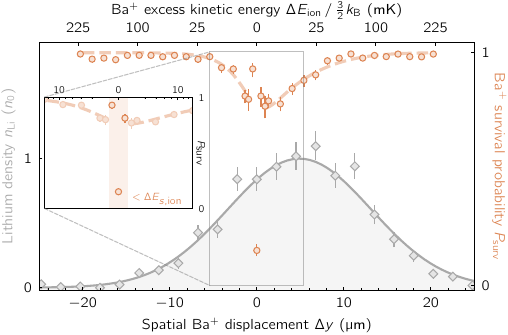}
    \caption{
        \textbf{Collision energy and density dependence of three-body recombination on an atom-ion Feshbach resonance}
        %\begin{itemize}
            %\item 
        We use a single electronically excited Ba$^+$ ion in the D$_{3/2}$ state to probe the \textit{in situ} density of the $^6$Li cloud
                (gray diamonds) in dependence on its displacement from the center of the rf trap $\Delta y$.
            %\item 
The density profile is in good agreement with a Gaussian (solid gray line) which is slightly offset from the center of the rf trap.
            %\item 
The survival probability of the Ba$^+$ ion prepared in the electronic ground state, $P_\text{surv}$, with the $B$-field tuned to
                resonance (\SI{321.9}{\gauss}) is probed for different excess kinetic energies $\Delta E_\text{ion}$ (orange data points).
            %\item 
At higher $\Delta E_\text{ion}$, the behavior of
                $P_\text{surv}$ is well described by a classical TBR model (dashed orange line).
            %\item 
When tuning to $\Delta E_\text{ion} <
                \SI{195}{\micro \kelvin}\times \frac{3}{2} k_\text{B}$, i.e.\ below the two-body $s$-wave limit $\Delta E_{s, \text{ion}}$
                (inset: dark orange markers, shaded region), we observe the emergence of resonant scattering, leading to a slight
                increase followed by a sharp decrease in $P_\text{surv}$. 
            %\item 
At the lowest $\Delta E_\text{ion}$, we experimentally observe a deviation of $21\sigma$ from the classical model.
            %\item 
The error bars indicate $1\sigma$ confidence intervals
                 and, where invisible, are smaller than the marker size.
        %\end{itemize}
            }
    \label{fig:fig2}
\end{figure}
%\begin{itemize}
    %\item 
To probe the density distribution of the Li ensemble $n(y)$, we perform ion loss spectroscopy with the ion
        prepared in the D$_{3/2}$ state and $t_\text{int}=\SI{40}{ms}$.
    %\item 
We displace the ion vertically from the center of the rf trap by applying
        $\mathcal{E}_\text{dc}$ and probe $P_\text{surv}$. 
    %\item 
This reveals a Gaussian-shaped density distribution $n(y)$ of width $\sigma=\SI{8.2(4)}{\micro m}$, corresponding
        to a FWHM of \SI{19.4(8)}{\micro\meter} (see Fig.~\ref{fig:fig2}). The cloud is offset from the center by $\mu =
        \SI{4.9(3)}{\micro m}$.
     %\item 
Interleaved with the density measurement, we prepare the ion in the S$_{1/2}$ state and probe
         $P_\text{surv}$ at $B=\SI{321.90(3)}{\gauss}$ after $t_\text{int}=\SI{200}{ms}$.
     %\item 
This allows us to simultaneously investigate the energy and density dependence of the TBR loss process.
    %\item 
At large $\Delta E_\text{ion} > \SI{100}{\milli\kelvin}\times\frac{3}{2} k_\text{B}$ and the corresponding
densities we
        find the TBR loss to be negligible.
    %\item 
Reducing the energy to $\Delta E_\text{ion} > \SI{200}{\micro\kelvin}\times\frac{3}{2}
        k_\text{B}$ leads to a decrease of $P_\text{surv}$.
    %\item 
In this regime, the experimental observation agrees with the established classical model, in which $\gamma_\text{l}$ is proportional to $E^{-3/4}$ and $n^2$ (see
Appendix~\ref{apdx:classical_tbr}) \cite{krukowEnergyScalingCold2016,
        perez-riosCommunicationClassicalThreshold2015}.
    %\item 
The feature, fitted with the classical model, has a FWHM of
        \SI{6.9(5)}{\micro\meter}, narrow compared to the density distribution of the atomic cloud.
        Further, the TBR loss is centered at zero displacement, albeit being slightly skewed due to the offset of the atomic cloud from the center of the
        rf trap.
        Both width and symmetry emphasize the importance of collision energy in the loss process.
     %\item 
Continuing to decrease $\Delta E_\text{ion} < \Delta E_{s, \text{ion}}$, we reveal a sharp modulation of $P_\text{surv}$,
        evidencing that the \SI{321.90(3)}{G} resonance begins to dominate the atom-ion
        interaction. 
    %\item 
At lowest $\Delta E_\text{ion}$, the experimental data deviate from the classical model by
        up to $21\sigma$.
    %\item 
We associate the deviation for $\Delta E_\text{ion}<\Delta E_{s,\text{ion}}$ with the transition
        to the quantum regime in which only the lowest
        partial waves contribute to the loss of the ion; that is, individual Feshbach resonances become
        pronounced.
%\end{itemize}

\section{Energy dependence of an $s$-wave resonance}

\begin{figure}[t]
    \centering
    \includegraphics{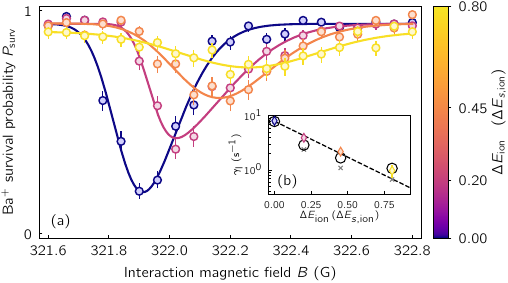}
    \caption{\textbf{Energy dependence of a low partial wave Feshbach resonance}\\
    %\begin{itemize}
        %\item 
        (a)
Ion-loss spectrum around the \SI{321.9}{G} resonance measured for different
            Ba$^+$ excess kinetic energies.
        %\item 
The blue (purple/orange/yellow) data points show $P_\text{surv}$ in dependence on the magnetic
            field $B$ for an excess kinetic energy of $\Delta E_\text{ion} \approx [0,0.20,0.45,0.80]\, \Delta E_{s,\text{ion}}$.
        %\item 
The temperature of the $^6$Li bath is \SI{700(50)}{\nano\kelvin} for all $\Delta
            E_\text{ion}$.
        %\item 
We observe that with increasing $\Delta E_\text{ion}$ the resonance is shifted towards higher magnetic
            fields, it decreases in amplitude and increases in width.
        %\item 
The solid lines are fits with a skewed Gaussian function.
    %\end{itemize}
(b)
    %\begin{itemize}
        %\item 
The peak loss rate  $\gamma_\text{l}$ is plotted as a function of $\Delta
            E_\text{ion}$ (diamonds).
        %\item 
The black dashed line is an exponential fit with a $1/e$--energy of
$0.33\,\Delta E_{s, \text{ion}}$.
        %\item 
The best quantum recombination model fit to the on-resonant loss rate is achieved for an
    $s$-wave resonance (black circles). For comparison, the best fit for a $p$-wave resonance is also
            shown (gray crosses).
        %\item 
The error bars represent $1\sigma$ confidence intervals and, where invisible, are smaller than the size of the markers.
    %\end{itemize}
}
    \label{fig:fig3}
\end{figure}
%\begin{itemize}
    %\item 
In the context of pronounced collision-energy effects below the $s$-wave barrier, we investigate the
\SI{321.90(3)}{G} resonance below $\Delta E_{s,\text{ion}}$ at four distinct 
        collision energies. 
    %\item 
We measure the Feshbach spectrum in the range \SIrange{321.6}{322.8}{\gauss} for 
$\Delta E_\text{ion} \approx [0, 0.2, 0.45, 0.8]\,\Delta E_{s, \text{ion}}$ and
        $t_\text{int}=\SI{200}{\milli\second}$.
    %\item 
In this regime, we observe a strong impact of the collision energy on the Feshbach
resonance (see Fig.~\ref{fig:fig3}): for higher energies the position of the resonance is shifted towards higher magnetic
        fields, its amplitude decreases and its width increases. An increase in energy to 
        $\Delta E_\text{ion}\approx 0.8\,\Delta E_{s,\text{ion}}$ ($\approx\SI{7}{\micro\kelvin}$ in
        the two-body COM) is enough to reduce the amplitude by more than 80\%. This is a clear
        indication of the dominant contribution of $\Delta E_\text{ion}$ to the COM collision energy.
    %\item 
We fit the series of resonance scans with skewed Gaussian functions, and
derive a linear dependence of the resonance position on $\Delta E_\text{ion}$ of \SI{2.4(2)}{\milli\gauss \per \micro\kelvin}.
    %\item 
Additionally, we extract the dependence of the peak loss rate $\gamma_\text{l}$ on
        $\Delta E_\text{ion}$ and find an exponential decrease with increasing
        energy (see Fig.~\ref{fig:fig3}b).
    %\item 
At first, this appears to contradict the threshold scaling laws, which predict an increase 
of the energy-dependent dimer coupling rate $\Gamma(k)$ for all partial waves. However,
in a
scenario, where the collision energies approach $E_s$ and the width of the collision energy
distribution is comparable to the natural linewidth of the resonance, a more intricate model is required
to understand the experimentally observed lineshapes and loss rates.
    %\item 
    We thus compare the scaling of $\gamma_\text{l}$ with the prediction of our quantum recombination model.
    %\item 

\section{Quantum recombination model}

    Our model is based on the two-step mechanism established for narrow resonances in
        neutral atoms~\cite{beaufilsFeshbachResonanceWave2009, foucheQuantitativeAnalysisLosses2019,
        liThreeBodyRecombinationNarrow2018}, but extended to take into account non-thermal
        energy distributions and beyond threshold effects. This is essential in the case of
        resonances that are narrow compared to the energy distribution and collision energies on the order of
        $E_s$.
    %\item 
In the model, a free atom-ion pair couples to a metastable atom-ion dimer state with partial-wave dependent 
        rate $\Gamma (k)$, and a secondary collision  with another atom at rate
        $\Gamma_\text{inel}=\mathcal{K}(k)n$ causes 
        inelastic recombination.
 The three-body cross section is given by
 \begin{align}
    \sigma(k)=\frac{\pi}{k^2}\frac{\Gamma(k)\mathcal{K}(k)n}{(\varepsilon-\varepsilon_{\rm res})^2/\hbar^2+(\Gamma(k)+\mathcal{K}(k)n)^2/4}
 \end{align}
    where $\Gamma(k)$ is the coupling rate with the dimer state, $\mathcal{K}(k)$ is the rate constant describing the inelastic atom-molecule collision, $\varepsilon_{\rm res}=\delta\mu(B-B_{\rm res})$ the resonance position, and $k$ is the wavevector corresponding to the collision energy in the three-body relative frame $\varepsilon$.
    %\item 
$\Gamma(k)$ can be separated into a product of the energy-independent short-range coupling $\Gamma_m$ and the
quantum defect function $C^{-2}(k)$ which provides the threshold behavior at low energy scaling $C^{-2}(k) \propto k^{2\ell+1}$ 
and approaches unity at large energies, providing a direct way to describe resonances far above the Wigner
threshold regime (see Appendix Fig.~\ref{fig:Apdx_dimer_formation}).
    %\item 
    For the inelastic rate constant $\mathcal{K}(k)$ we employ the quantum defect model, assuming that the
    collision between the atom and the two-body complex is universal, i.e.\ the probability of reaction at
    short range approaches unity. The resulting energy-dependent reaction rate does not depend on any free
    parameters, although this assumption can be relaxed~\cite{gaoQuantumLangevinModel2011, jachymskiQuantumTheoryReactive2013}.
    %\item 
    At very low collision energies, i.e.\ for $\Gamma(k)\ll \mathcal{K}(k)n$, the loss rate on
    resonance follows $\gamma_\text{l}=\sigma(k)n v \propto k^{2\ell}$. On the other hand above threshold,
where $\Gamma(k)=\Gamma_m$, the loss rate for resonances from any open channel decreases with
$\gamma_\text{l}\propto k^{-1}$. Another important case is that of $\Gamma(k)\gg\mathcal{K}(k)n$ at collision
energies well below the height of the partial wave barrier. Here we find that $\gamma_\text{l}\propto
k^{-2(\ell+ 1)}$. These effects can, for $\ell>0$, account for a maximum of $\gamma_\text{l}$ that can lie
well below the respective threshold.
For comparison to the experimental data the rates $\Gamma(k)$ and $\Gamma_\text{inel}(k) = \mathcal{K}(k)n$ have to be averaged over the proper
velocity distribution, taking into account that the sample 
is non-thermal (the ion and the atoms follow different energy distributions and we are interested in the energy in the three-body relative frame).
We extract the respective distributions from MD simulations (see
Appendix~\ref{apdx:atom_ion_simulation}). Note that in our model, this averaging leads to a shift of the
resonance, which, based on the scaling laws mentioned above, depends on $\ell$.
    %\item 
At this point the model neglects scattering in other partial waves, spin-orbit coupling effects as well as Stark shifts from the lasers and the rf field.
        
    %\item 
Comparing experimental and theoretical results we find the best agreement for an open channel $s$-wave
resonance (see Fig.~\ref{fig:fig3}b) with a weighted sum of squared residuals of $\chi^2\approx14$. In this
case we find a coupling rate of $\Gamma_m = 2\pi\times\SI{768(11)}{\kilo\hertz}$ and a relative magnetic moment of $\delta\mu = \SI{1.86(4)}{\mega\hertz\per G}$.
    %\item 
When modeling the \SI{321.9}{G} resonance with a $p$-wave resonance, we find worse agreement with $\chi^2\approx 41$.
        In contrast, according to our model, resonances attributed to higher partial wave contributions show a
        qualitatively different behavior. With increasing $\Delta E_\text{ion}$, they exhibit a significant
        increase in amplitude in the range below $\Delta E_{s, \text{ion}}$, in contradiction to our
        experimental findings.
%\end{itemize}

\section{A higher-partial-wave resonance}

\begin{figure}[t]
    \centering
    \includegraphics{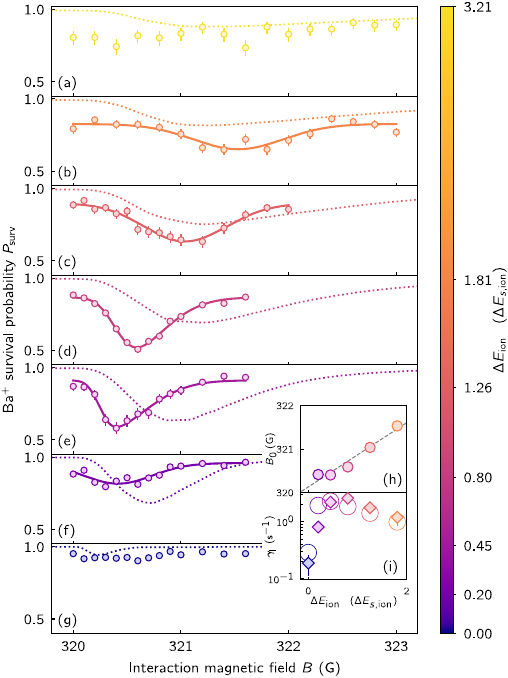}
    \caption{\textbf{Emergence and fading of a higher-partial-wave Feshbach resonance}\\
        %\begin{itemize}
            %\item 
        (a)--(g)
Ion-loss spectrum between \SI{320}{\gauss} and \SI{323}{\gauss} for increasing excess kinetic
                energies $\Delta E_\text{ion}$ between 0 and
                $3.21\,\Delta E_{s,\text{ion}}$ (from bottom to top).
            %\item 
Increasing the excess kinetic energy gives rise to a Feshbach resonance 
                that emerges, peaks in amplitude, and disappears in the background at higher collision energies.
            %\item 
Along with the effects on the resonance we observe a decrease in $P_\text{surv}$ on background 
                with increasing collision energy.
            %\item 
The solid lines are fits with a skewed Gaussian function.
            %\item 
The error bars indicate $1\sigma$ confidence intervals.
        %\end{itemize}
        %\begin{itemize}
            %\item 
The dotted lines show the result of the quantum recombination model.
            %\item 
An open-channel $f$-wave resonance is modeled for
                the same seven $\Delta E_\text{ion}$ chosen experimentally.
            %\item 
    %\end{itemize}
                (h) -- (i) The resonance position $B_0$ (filled circles) and peak loss rate
                    $\gamma_\text{l}$ (filled diamonds) obtained from the fits are shown in dependence on
                    $\Delta E_\text{ion}$. The grey dashed line is a linear fit with a slope of
                $\SI{4.4(4)}{\milli G \per \micro\kelvin}$. We compare $\gamma_\text{l}$ to the prediction of
            the quantum recombination model (open circles) and find qualitative agreement.
}
    \label{fig:fig4}
\end{figure}

%\begin{itemize}
    %\item 
To elucidate the role of higher partial wave channels experimentally, we perform ion loss spectroscopy in the
neighboring range of \SI{319}{\gauss}
        to \SI{323}{\gauss} with $\Delta E_\text{ion} \approx [0, 0.2, 0.45, 0.8, 1.26, 1.81,
        3.21]\,\Delta E_{s,\text{ion}}$ (see Fig.~\ref{fig:fig4}).
    %\item 
At $\Delta E_\text{ion}=0$ we observe no significant signal between \SI{320}{\gauss} and
        \SI{321.8}{\gauss}.
    %\item 
With increasing energy, a resonance emerges at \SI{320.41(3)}{G}. It peaks in amplitude at
        $B=\SI{320.59(3)}{\gauss}$ and $\Delta E_\text{ion}\approx0.8\,\Delta E_{s, \text{ion}}$.
    %\item 
Its position is further shifted to $B=\SI{321.53(3)}{\gauss}$ at $\Delta
E_\text{ion}\approx 1.8\,\Delta E_{s, \text{ion}}$ before
        its signature fades at higher energies.
    %\item 
We observe a dependence of the resonance position on $\Delta E_\text{ion}$ of 
        \SI{4.4(4)}{\milli\gauss\per\micro\kelvin}, around
        twice as large as that of the \SI{321.9}{\gauss} resonance.
    %%\item 
Both the peak amplitude at higher $\Delta E_\text{ion}$ and larger shift indicate that the resonance stems
from a higher open-channel partial wave.
    %    partial wave.
    %\item 
 Assuming a higher-partial-wave resonance, we compare the experimental results to an $f$-wave resonance
 with a relative magnetic moment of \SI{0.35}{\mega\hertz\per\gauss}
 that we describe by the quantum recombination model. Choosing the same seven values for $\Delta E_\text{ion}$, we find that it is very weak at low collision energies but peaks at $\Delta
 E_\text{ion} = 0.45\,\Delta E_{s, \text{ion}}$.
    %\item 
At higher energies, the modeled resonance is further  
        broadened and becomes less pronounced.
        Note that, unlike for the \SI{321.9}{G} resonance, the modeled $f$-wave resonance is not the
            result of an
        optimization. We see comparable behavior when modeling an open-channel $d$-wave resonance, however
    $s$- and $p$-wave resonances scale qualitatively different.
    %\item 
    Based on our experimental findings and their qualitative agreement with the quantum
    recombination model, we conclude that
        the observed resonance stems from a higher partial wave channel.
    %\item 
Furthermore, we observe a decrease of $P_\text{surv}$ independent of $B$ when increasing $\Delta
        E_\text{ion}$, evidencing the contribution of higher partial-wave channels to background loss.
        Note that the quantum recombination model, at this point, does not take the effect of the background
        scattering length into account.  
%\end{itemize}

\section{Conclusion and Outlook}
%\begin{itemize}
    %\item 
In this Article, we have studied the energy dependence of atom-ion three-body recombination
        across several orders of magnitude from tens of \si{\milli\kelvin} to below $E_{s}$  witnessing the
        transition from the classical to the quantum regime.
    %\item 
In the many partial-wave regime ($\Delta E_\text{ion} > \SI{10}{\milli\kelvin}$)
        our results can be described by a classical atom-atom-ion recombination model.
    %\item 
In the few partial-wave regime, where Feshbach resonances dominate the scattering process,
        we identified and characterized two individually resolved resonances with different collision energy scaling.
    %\item 
While one decreases in amplitude with increasing energy, the other initially increases and peaks at a distinct energy.
    %\item 
This implies that the resonances originate in two different open-channel partial waves.
    %\item 
Based on the comparison with a beyond-threshold quantum recombination model we conclude that the former
         is an $s$-wave while the latter is a higher partial-wave resonance.
    %\item 
         Both resonances show a strong dependence on energy below the two-body $s$-wave limit, underlining
         both the importance of reaching low collision energies and the significance of higher partial wave
         resonances for atom-ion interaction in the ultracold regime.

We expect our findings to advance the understanding of the Li-Ba$^+$ spectrum and more generally of atom-ion
        interaction at ultracold temperatures. The experimental assignment of open-channel partial waves
        enables the theoretical exploration
        of the role of close-range atom-ion interactions such as second-order spin-orbit coupling or
        spin-spin interaction. Here the experimental assignment of the open-channel partial wave could allow
        the direct observation of orbital angular momentum changing collisions.
    %\item 
Furthermore, we introduce ion excess kinetic energy as an additional parameter for fast control over
         atom-ion scattering that can be applied even when rapid changes in the magnetic field are
         disadvantageous. At a constant magnetic field, the atom-ion system can be tuned to resonance with a
         higher partial wave resonance by applying an
        electric displacement field -- in principle on a microsecond timescale.
    %\item 

        However, we have also shown that at the lowest collision energies currently accessible to hybrid trap setups, non-zero
        angular momentum resonances still play a significant role.
    %\item 
        Combining
        atoms and an ion in an optical dipole trap could allow to avoid any micromotion and reach even lower
        collision energies \cite{schaetzTrappingIonsAtoms2017, lambrechtLongLifetimesEffective2017, schmidtOpticalTrapsSympathetic2020}. This would allow to better distinguish $s$- and $p$-wave resonances and provide a path
        towards the exploration of potentially even more temperature-sensitive many-body
        complexes.
    %\item 
        This approach could further allow to experimentally access species combinations of 
        generic mass ratios and extend the single-particle control to the neutral atoms using tightly
        focussed optical tweezers \cite{endresAtombyatomAssemblyDefectfree2016}.
    %\item 
Atom-ion systems offer a unique possibility among ultracold systems as the kinetic energy of the atoms and
        the ion can be tuned individually. However, changing the temperature of the bath comes along with a change in trap laser intensity. 
        The respective impact of  
        bath temperature and trap laser intensity could be disentangled by adiabatic decompression of the atom
        trap, resulting in different kinetic energy distributions at similar laser intensities.
        In this matter, also broader resonances could be of interest, to test the model in a regime where the collision energy
        distribution is narrow compared to the linewidth of the resonance.
    %\item 
Another interesting property of the bath is its quantum statistics. Admixing the spin-polarized
        Fermi gas of the experiments presented here with a second spin component
        could provide insight into the
        role of Pauli exclusion in atom-atom-ion recombination.
    %\item 
This might help to validate whether the two-step model, assuming a resonant two-body and
        subsequent inelastic loss process, adequately describes atom-atom-ion recombination
        \cite{lecomteLossFeaturesUltracold2024}. 
    %%\item 
    %  our and other recent results emphasize that the properties of atom-ion collisions -- even at the
    %    lowest achievable temperatures -- are far from being understood.
%\end{itemize}

%
\clearpage

\appendix

\section{Experimental setup}
    %\item 
%\item 
We operate our rf trap at $\Omega_\text{rf}\approx 2\pi\times\SI{1.43}{\mega\hertz}$ creating a time-averaged
    pseudo-potential with secular frequencies $\omega^\text{Ba}_{1,2,3} = 2\pi \times [\num{66.9}, \num{64.9}, \num{7.2}]\,\si{\kilo\hertz}$.
%\end{itemize}
The generic experimental protocol for ion-loss spectroscopy is as follows:
%\begin{itemize}
    %\item 
        We initially load Ba$^+$ ions via ablation loading and, if necessary, isolate a single ion via isotope
        selective excitation and optical trapping \cite{weckesserTrappingShapingIsolating2021,
        schmidtMassselectiveRemovalIons2020}.
        We then Doppler-cool it and prepare it via optical pumping, either in the electronic ground state S$_{1/2}$ or
        the metastable excited D$_{3/2}$ state. At this point, we do not spin-polarize the ion and,
            thus, expect
        a mixture of its Zeeman states. We compensate stray electric fields with an accuracy of
        $\mathcal{E}_\text{stray}\approx \SI{3}{\milli\volt\per\meter}$. Then we apply dc control voltages to axially
        shift the ion out of the center of the rf trap, where we subsequently load a cloud of $^6$Li
        atoms.
    %\item 
We evaporatively cool the Li atoms down to temperatures as 
low as \SI{700(50)}{\nano\kelvin} and spin-polarize them in their lowest hyperfine state
$\ket{1}_\text{Li}=\ket{m_s=-1/2, m_I=1}$. 
In this hyperfine state, spin exchange collisions that can occur when Li is prepared in the second lowest hyperfine state
\begin{align*}
    |m_s =&-1/2,\, m_I=0\rangle_\text{Li} + \ket{m_s=1/2}_\text{Ba} \\ \qquad &\rightarrow \ket{m_s=-1/2,
    m_I=1}_\text{Li} + \ket{m_s=-1/2}_\text{Ba},
\end{align*}
 are forbidden and cannot lead to heating effects.
        Typically, we
        prepare $N\approx \num{1.7e4}$ atoms at a density of $n\approx \SI{5e11}{\per\cubic\centi\meter}$.
    %\item 
The ion is then immersed into the bath of ultracold atoms for an
        interaction duration $t_\text{int}$. During $t_\text{int}$, we apply a magnetic field $B$ and,
        optionally, a dc
        displacement field $\mathcal{E}_\text{dc}$ along the $y$-direction to control the collision energy (see below).
    %\item 
After $t_\text{int}$, we interrogate the ion product state via fluorescence detection to determine
whether it survived the interaction , i.e.\ it remained cold and in its electronic ground state (see
Appendix~\ref{apdx:ion_state}).
    %\item 
We repeat the protocol to derive the ion survival probability $P_\text{surv}$ for a given set
        of experimental parameters, as well as the related statistical uncertainty based on Wilson score intervals.

\section{Overlap of ion and atomic cloud}
%\begin{itemize}
    %\item 
The atomic cloud is trapped in a \SI{1064}{nm} crossed optical-dipole trap (xODT), consisting of two
        beams that intersect at an angle of \SI{14}{\degree} and lie in a plane that forms an angle of
        \SI{31}{\degree} with respect to the axial direction of the rf trap.
    %\item 
We use piezo mirrors and the AC-Stark shift on the ion to align the trapping beams with the center
        of the rf trap.
    %\item 
During ion-loss spectroscopy experiments, the whole apparatus reaches a steady state, leading to a shift of the
        xODT position that we observe by absorption imaging. The magnitude of this shift is consistent with
the offset observed in Fig.~\ref{fig:fig2}.
%\end{itemize}

\section{Ion state preparation and readout}
\label{apdx:ion_state}
%\begin{itemize}
    %\item 
The ion is initially Doppler cooled using the S$_{1/2}$ to P$_{1/2}$ transition, in combination with
        a repumping laser from the D$_{3/2}$ state. To prepare the ion in the S$_{1/2}$ (D$_{3/2}$) state, we switch off the cooling (repumping) laser \SI{50}{\milli\second} before the end of the Doppler cooling phase.
        At this point we do not spin polarize the ion in a specific Zeeman state.
    %\item 
In addition to Doppler cooling, limited to $\approx \SI{360}{\micro\kelvin}$, the ion is
        sympathetically cooled in the ultracold atomic bath to lower temperatures. From simulations, we obtain
        a lower limit for the median ion kinetic energy of $\SI{2.2}{\micro\kelvin\times}k_\text{B}$. This
        is consistent with our observation of a broadening of the 321.9G resonance, when we
        increase the temperature of the atomic bath from \SI{700(50)}{\nano\kelvin} to \SI{2.8(2)}{\micro\kelvin}. From the 
        experimental results presented in this article, we also conclude an upper limit of
        $\approx\SI{30}{\micro\kelvin\times}k_\text{B}$, as increasing $\Delta E_\text{ion}$ by this amount, gives
        rise to significant changes in the resonance shape.
    %\item 
After the interaction, we interrogate the ion product state based on fluorescence detection of the
        ion. First, we only switch on a near-detuned cooling laser and the repumping laser from the D$_{3/2}$ state, then we successively add a far-detuned
        cooling laser and a repumping laser from the D$_{5/2}$ state. In this way, we distinguish between a direct
        detection (``survival''), a heated ion (``hot''), an ion in the metastable D$_{5/2}$ state and ion
        loss from the trap.
        We found no significant effect of either the magnetic field or the collision energy on the
        distribution of the three inelastic events.
    %\item 
        Thus, in this article, we refer to the probability of a survival event as $P_\text{surv}$ and, for the sake
        of readability, define any signature of an inelastic process as loss, i.e. $P_\text{loss} =
        1-P_\text{surv}$.
    %\item 
We performed independent measurements to verify that $P_\text{surv}$ follows an
        exponential decay with $t_\text{int}$ for both the inelastic two-body loss and TBR. This allows us to extract the loss rate 
        $\gamma_\text{l} = -\frac{\ln P_\text{surv}}{t_\text{int}}$. Note that this
is different from the case of neutral-atom experiments, in which the density dependence of different loss
        processes leads to different superexponential scalings. 
    %\item 
        The experimental data presented in Fig.~\ref{fig:fig2} and \ref{fig:fig3} was recorded in a fully interleaved fashion. The spectra presented in
        Fig.~\ref{fig:fig4} were partially recorded in individual runs.
%\end{itemize}

\section{Magnetic field calibration}
%\begin{itemize}
    %\item 
We calibrate the magnetic field by performing rf-spectroscopy on the Li
            $\ket{1} \rightarrow \ket{2}$ hyperfine transition. 
            Long-term measurements reveal that the field is stable within $\sigma_{B, \text{stat}} = \SI{24}{\milli\gauss}$
            over the course of 12 hours. The systematic uncertainty of the calibration for the entire range of
            B is $\sigma_{B,\text{sys}}=\SI{80}{\milli\gauss}$.
    %\item 
\section{Tuning the ion excess kinetic energy}
\label{apdx:atom_ion_simulation}
%\begin{itemize}
    %\item 
We run MD simulations to find the kinetic energy scaling of the ion with respect to the applied displacement field
        $\mathcal{E}_\text{dc}$ \cite{furstProspectsReachingQuantum2018, trimbyBufferGasCooling2022b}. The simulations are performed in three
        spatial dimensions and include the radial rf fields, as well as the long-range attractive $C_4$ and
        short-range repulsive
        $C_6$ atom-ion interaction potential. At this point, we neglect the contribution of phase micromotion. 
        We simulate a large number of scattering events between the ion and a single atom.  
        For each scattering event, the atom is initialized on a sphere with radius $r_\text{init}$ around the ion. 
        The atomic kinetic energies are sampled from a thermal distribution. The ion is initially at rest
        and, over time and averaging over many trajectories, reaches a steady-state median kinetic energy. 
        We perform these simulations for different displacement fields $\mathcal{E}_\text{dc}$ and obtain the scaling factor
        \begin{align}
            \alpha_\text{sim} \approx \SI{10(1)}{\milli\kelvin \per (\volt\per\meter)^2}
        \end{align}
        corresponding to $\approx \SI{420(40)}{\micro\kelvin \per (\volt\per\meter)^2}$ in the two-body COM
        (see Fig.~\ref{fig:energy_scaling}). 
    %\item 
        Note that we employ the median kinetic energy, as the characteristic power law distribution of the ion's
        kinetic energy after equilibrating with the atomic bath does not have a mean value due to the strong influence of
        high-energy outliers. We observe this in our simulations in the form of non-converging mean energies.
    %\item 
Independently, we can estimate $\alpha$ for the isolated ion based on the rf trap parameters 
        \cite{berkelandMinimizationIonMicromotion1998} as
        \begin{align}
            \label{eq:berkeland_e_mm}
            \alpha_{\text{theo},i} = \frac{4}{m} \left(\frac{e q_i}{(2a_i + q_i^2) \Omega_\text{rf}} \right)^2,
        \end{align}
    with $i$ indicating the spatial direction, the trap parameters $a_i$ and $q_i$, the mass of the ion $m$,
    the elementary charge $e$ and the angular drive frequency of the rf trap $\Omega_\text{rf}$. For our trap configuration
    and displacement in the y-direction we obtain
    \begin{align}
       \alpha_{\text{theo},y} \approx \SI{16.4(10)}{\milli\kelvin \per (\volt\per\meter)^2}. 
    \end{align}

    %\item 
We thus find a \SI{60}{\percent} larger value compared to simulating the full atom-ion dynamics. When we
simulate the ion without interaction with the atoms, we obtain \SI{14(2)}{\milli\kelvin \per
(\volt\per\meter)^2}. Thus we attribute the difference to the effect of atom-ion collisions as well as using
the median, instead of the mean energy.

    %\item 
Another feature of the kinetic energy distribution of the ion is a strong anisotropy (see 
Fig.~\ref{fig:energy_scaling} b and c). As
        $\mathcal{E}_\text{dc}$ is applied in the radial direction, the kinetic energy distribution in the radial plane
        has a strong nonthermal component. In an ideal trap, the ion is only heated axially 
        by collisional redistribution of the radial kinetic energy. This results in a kinetic energy
        distribution much closer to that of a thermal ensemble. We use simulated 3D ion velocity
        distributions when modeling the behavior of resonances with the quantum recombination model.
 %\end{itemize}

\section{Additional confidence tests}
\begin{figure}[t]
    \centering
    \includegraphics[]{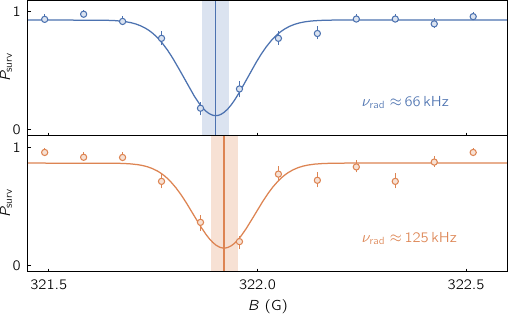}
    \caption{\textbf{Role of the rf-confinement for an atom-ion Feshbach resonance.} We show the ion loss
    spectrum around the \SI{321.9}{G} resonance for two different rf-confinements of
    $\omega\approx2\pi\times\SI{66}{kHz}$ (blue data) and $\omega\approx2\pi\times\SI{125}{kHz}$ with no
    displacement field $\mathcal{E}_\text{dc}$ applied. A fit with a Gaussian function (solid lines) is used to
    determine the resonance position (vertical lines, the uncertainty is indicated by the shaded regions). The
    measurement was performed in an interleaved fashion. The errorbars indicate 1$\sigma$ confidence intervals
and, where invisible are smaller than the marker size.}
    \label{fig:apdx_different_rf_confinements}
\end{figure}
%\begin{itemize}
    %\item 
To ensure that our observations can be attributed to collision-energy effects, in addition to the
        evidence already presented in the main text, we perform the following confidence tests:
    %\item 
First, we use a different method to tune
the collision energy. 
We vary the atomic bath temperature between \SI{700}{\nano\kelvin} and
\SI{11}{\micro\kelvin} and measure the width of the \SI{321.9}{\gauss} resonance, finding that it increases with a
slope of \SI{50(5)}{\milli\gauss\per\micro\kelvin}, in reasonable agreement with \SI{42(4)}{\milli\gauss\per\micro\kelvin}
when varying $\Delta E_\text{ion}$. This serves as a strong indication that the observed effects are indeed
caused by the collision energy. In addition, this validates that the energies relevant for the \SI{321.9}{G}
resonance are on the order of magnitude of a few \si{\micro\kelvin}. 
    %\item 
However, other effects, such as an overall decrease of $P_\text{surv}$ independent of $B$, likely associated with the increase in the intensity of the xODT beams, make it difficult to directly compare this method with the ion excess kinetic energy
        method. The impact of the trap light, for example light-assisted losses and an AC-Stark shift of the
        resonance, has also been reported elsewhere
        \cite{khlebnikovRandomChaoticStatistic2019} and is currently under
        investigation for Li-Ba$^+$.
    %\item 
        Second, we increase the radial confinement provided by the rf fields by a factor of two (see
        Fig.~\ref{fig:apdx_different_rf_confinements}). At $\Delta
        E_\text{ion}=0$, we do not find any statistically significant differences in the shape, position, or amplitude of the \SI{321.9}{\gauss} resonance. From this we conclude that trap-induced bound states, as reported in \cite{hirzlerTrapAssistedComplexesCold2023,pinkasTrapassistedFormationAtom2023}, do not significantly
affect the resonance, likely due the high mass imbalance and low rf-trap frequencies in our setup.
    %\item 
Increasing $\Delta E_\text{ion}$ at higher confinement, we find that applying approximately twice the
        displacement field results in a similar shift and decrease in amplitude as presented in
        Fig.~\ref{fig:fig3}. This is in agreement with Eq.~\ref{eq:berkeland_e_mm}.
%\end{itemize}

\section{Classical description of three-body recombination loss}
\label{apdx:classical_tbr}
%\begin{itemize}
    %\item 
In an independent measurement we
        confirm that, on a Feshbach resonance, the loss rate $\gamma_\text{l}$ is proportional to
        $n^2$\cite{weckesserObservationFeshbachResonances2021}.
    %\item 
It was further shown in experiment and theory that in the classical
            regime the atom-ion TBR rate scales with $E^{-3/4}$
            \cite{perez-riosCommunicationClassicalThreshold2015, krukowEnergyScalingCold2016}.
    %\item 
To model the energy at which the survival probability no longer follows a
classical scaling, we introduce a minimum energy $E_0$. The loss rate is thus 
\begin{align}
    \gamma_\text{l}  = k_3 n^2 (E_0 + \Delta E_\text{ion})^{-3/4}.
\end{align}
    %\item 
        We find a good fit with our experimental data taken at $\Delta E_\text{ion}>\Delta E_{s, \text{ion}}$ for $E_0  =
        \SI{4.5(8)}{\milli\kelvin}\times\frac{3}{2} k_\text{B}$ (corresponding to \SI{190(30)}{\micro\kelvin} in the two-body COM frame)
        and a $k_3$-parameter of $\SI{2.31(23)e-23}{cm^{-6}\second^{-1}}$ at $\SI{10}{\milli\kelvin}$.
%\end{itemize}

%
\begin{figure}[t]
    \centering
    \includegraphics[]{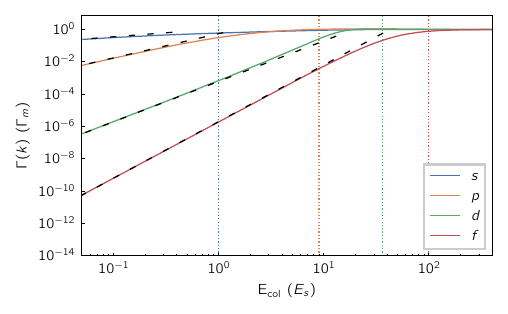}
    \caption{\textbf{Energy scaling of the dimer coupling rate:} The coupling rate
    $\Gamma(k)$ with the dimer state is plotted in dependence on the collision energy for the lowest four
    partial waves. The black dashed lines indicate threshold scaling laws $k^{2\ell+1}$, which are a
    valid description at low energies. At energies above the height of the partial wave barrier $E_\ell$
(vertical dotted lines), $\Gamma(k)$ reaches unity.}
    \label{fig:Apdx_dimer_formation}
\end{figure}

\clearpage
\section{Acknowledgments}
This project has received funding from the European Research Council (ERC) under the European Union’s Horizon
2020 research and innovation programme (grant number 648330), the Deutsche Forschungsgemeinschaft (DFG, grant
number SCHA 973/9-1-3017959) and the Georg H. Endress Foundation. F.T., J.S., D.v.S. end T.S. acknowledge financial support from the DFG via the RTG
DYNCAM 2717. W.W. acknowledges financial support from the QUSTEC programme, funded by the
European Union's Horizon 2020 research and innovation programme under the Marie Skłodowska-Curie (grant number
847471). P.W. gratefully acknowledges financial support from the Studienstiftung des deutschen Volkes. K.J.
was supported by the Polish National Agency for Academic Exchange (NAWA) via the Polish Returns 2019
programme. T.W. and T.S. acknowledge financial support from the Georg H. Endress foundation.
We thank Ulrich Warring, Deviprasath Palani, Dietrich Leibfried, Jesús Pérez-Ríos, Masato Morita and Michał Tomza for
fruitfull discussions.

\section{Authorship}
F.T. and J.S. carried out the experiments and analyzed the data with support from T.W. and under supervision
of T.S. F.T., J.S., D.v.S., W.W.
and P.W. built the experiment. F.T. performed the MD
simulations. K.J. performed the theoretical calculations. F.T. and T.S. prepared the manuscript, with
contributions from all authors. All authors participated in the interpretation of the data.

\section{Data availability}
The data presented in this paper is available from the corresponding author upon request.

\section{Code availability}
The \texttt{julia} code used to simulate the ion dynamics in the atomic bath is available upon request. The code used to analyze the raw data is available upon request. The code used to simulate the quantum recombination model is available upon request.

\bibliography{zotero_bib}
%\printbibliography

\end{document}